\documentclass[twocolumn,preprintnumbers,superscriptaddress,nofootinbib,aps,prd,floatfix]{revtex4}

\usepackage{amsmath,amssymb}
\usepackage{graphicx} 
\usepackage{epstopdf} 
\usepackage{slashed}
\usepackage{subfigure}
\usepackage{xcolor}
\usepackage{multirow}
\usepackage{hyperref}

\usepackage{footmisc}

\usepackage{array}
\newcolumntype{L}[1]{>{\raggedright\let\newline\\\arraybackslash\hspace{0pt}}m{#1}}
\newcolumntype{C}[1]{>{\centering\let\newline\\\arraybackslash\hspace{0pt}}m{#1}}
\newcolumntype{R}[1]{>{\raggedleft\let\newline\\\arraybackslash\hspace{0pt}}m{#1}}

\newcommand{\be}{\begin{eqnarray*}}
\newcommand{\ee}{\end{eqnarray*}}

\newcommand{\bee}{\begin{eqnarray}}
\newcommand{\eee}{\end{eqnarray}}
\newcommand{\beeq}{\begin{equation}}
\newcommand{\eeeq}{\end{equation}}

\usepackage{bbold}

\begin{document}

\title{Anomalous neutral gauge boson interactions and simplified models}
\begin{abstract}
Trilinear $Z$ boson interactions are sensitive probes both of new sources of $CP$ violation in physics Beyond the Standard Model and of new particle thresholds. Measurements of trilinear $Z$ interactions are typically interpreted in the frameworks of anomalous couplings and effective field theory, both of which require care in interpretation. To obtain a quantitative picture of the power of these measurements when interpreted in a TeV-scale context, we investigate the anatomy of $ZZZ$ interactions and consider two minimal and perturbative simplified models which induce such interactions through new scalar and fermion loops at the weak scale, focusing on $ZZ$ and vector boson fusion-induced $Zjj$ production at the LHC and $ZZ$ production at a future $e^+ e^-$ collider. We show that both threshold and non-threshold effects often are small compared to the sensitivity of the LHC, while the increased sensitivity of a future lepton collider should allow us to constrain such scenarios through associated electroweak precision effects complementary to direct searches at hadron colliders.
\end{abstract}
%
%
\author{Tyler Corbett} \email{tyler.corbett@unimelb.edu.au}
\affiliation{ARC Centre of Excellence for Particle Physics at the Terascale, School of Physics, The University of Melbourne, Victoria, Australia\\[0.1cm]}
\author{Matthew J. Dolan} \email{dolan@unimelb.edu.au}
\affiliation{ARC Centre of Excellence for Particle Physics at the Terascale, School of Physics, The University of Melbourne, Victoria, Australia\\[0.1cm]}
\author{Christoph Englert} \email{christoph.englert@glasgow.ac.uk}
\affiliation{SUPA, School of Physics and Astronomy, University of
  Glasgow, Glasgow G12 8QQ, United Kingdom\\[0.1cm]}
\author{Karl Nordstr\"om} \email{k.nordstrom@nikhef.nl}
\affiliation{SUPA, School of Physics and Astronomy, University of
  Glasgow, Glasgow G12 8QQ, United Kingdom\\[0.1cm]}
\affiliation{Nikhef, Science Park 105, NL-1098 XG Amsterdam, The Netherlands\\[0.1cm]}

\preprint{Nikhef 2017-054}

\maketitle

\section{Introduction}
The discovery of the Higgs boson at the Large Hadron Collider has not revealed conclusive hints towards new phenomena beyond the Standard Model (BSM) and the
nature of the mechanism of electroweak symmetry breaking remains
elusive. Taking this lack of BSM physics at face value, the high energy physics community has moved towards studying the Standard Model (SM) as a low-energy effective field theory (EFT),  paving the way towards comprehensive data analyses in a dimension six extended SM-EFT framework~\cite{Henning:2014wua,Brivio:2017vri}.
As the EFT parameterization of our ignorance towards BSM physics includes \emph{all} possible UV completions of the SM, this suggests that collider processes can receive corrections from multiple and competing EFT terms which can potentially introduce issues when calculating these processes perturbatively.

$Z$-boson pair production~\cite{PhysRevD.43.3626,MELE1991409,Biedermann:2016yvs,Biedermann:2016lvg,Heinrich:2017bvg} and $Z$+2~jet production via weak boson fusion~\cite{Chehime:1992ub,Rainwater:1996ud,Oleari:2003tc,Denner:2013fca,Schissler:2013nga,Denner:2014ina,Blumenschein:2018gtm} are standard candles that inform both SM and BSM interpretations of LHC measurements. In particular, they are sensitive to new particle thresholds as well as the presence of a high scale-induced $ZZZ$ interaction, which is a sign of $CP$-violation beyond that within the SM. The presence of a new source of $CP$ violation is required to explain the baryon asymmetry of the universe~\cite{Rubakov:1996vz,Riotto:1999yt,Morrissey:2012db}, and links the collider phenomenology of the $ZZZ$ vertex to baryogenesis. Interactions which induce such couplings exist in a range of models~\cite{Fairbairn:2013xaa,Dorsch:2013wja,Inoue:2014nva,Inoue:2015pza,Dorsch:2016nrg,Voigt:2017vfz}. Accordingly, such trilinear gauge couplings have been searched for by the ATLAS, CMS, and L3 Collaborations~\cite{Achard:2003hg,CMS:2014xja,Khachatryan:2015pba,Aaboud:2016urj,CMS:2017ruh,Aaboud:2017rwm,Sirunyan:2017zjc}, which have placed constraints on their existence using an approach based on anomalous couplings. Additionally the L3 Collaboration has searched for direct signatures of models which can provide an explicit realization of such a vertex~\cite{Achard:2001qw}.

The anomalous coupling approach is not ideal from a theoretical perspective. It is not gauge invariant from an electroweak point-of-view, and leads to violation of unitarity bounds at LHC energies. This unitarity violation is often overcome through the use of momentum dependent form factors (as used by ATLAS in~\cite{Aaboud:2016urj} for instance, following~\cite{Hagiwara:1986vm,Baur:1987mt,Baur:1988qt,Baur:2000ae}). However, these form factors are themselves not well motivated~\cite{Degrande:2012wf}. A more robust approach is to use effective field theory (EFT), by adding gauge-invariant higher-dimensional operators to the SM Lagrangian. Indeed, there have been a number of recent studies of the phenomenology of trilinear gauge couplings from this perspective~\cite{Falkowski:2016cxu,Azatov:2017kzw}. Perturbative unitarity violation remains a possibility in such an approach. However, as matching to concrete UV scenarios becomes possible beyond the limitations of form factors, violating perturbative unitarity bounds translates into a non-existing constraint for perturbative UV scenarios at the high scale, and therefore does not limit the use of EFT as a mediator between theories at different scales.

Effective field theories work best when there is a clear hierarchy between the energy scales being probed experimentally and the fields which have been integrated out. This is not always the case for the parameter space of interest in UV complete models. On the one hand, theories with new sources of $CP$ violation in the context of electroweak baryogenesis generally require new fields with masses close to the electroweak scale~(see e.g.~\cite{Fairbairn:2013xaa,Dorsch:2013wja,Inoue:2014nva,Inoue:2015pza,Dorsch:2016nrg}). On the other hand, for extended fermion sectors, as predicted for instance in scenarios of partial compositeness~\cite{Kaplan:1991dc,Contino:2006nn}, $ZZZ$ interactions can be sourced at one-loop by non-diagonal $Z$ couplings to top-quark or lepton partners. Both cases can imply marked changes in collider observables, as amplitudes become imaginary when virtual particles are able to go on-shell.

It is the purpose of this paper to bridge the gap between anomalous couplings/EFTs and UV complete theories by studying (gauge-invariant) simplified models~\cite{Alves:2011wf}. These have proved of great utility in searches for supersymmetry~\cite{Alves:2011wf,Cohen:2013xda} and dark matter production~\cite{Buchmueller:2013dya,Buchmueller:2014yoa,Abdallah:2015ter} at the LHC. This allows us to gauge the reported constraints from LHC precision measurements in a more realistic context relevant for TeV scale physics.

If a simplified model is also renormalisable, it opens up the possibility to correlate oblique electroweak precision~\cite{Peskin:1990zt,Peskin:1991sw,Barbieri:2004qk} measurements (for earlier EFT-related work see~\cite{Hagiwara:1992eh,Hagiwara:1993ck,Alam:1997nk}) with the potential sensitivity of $ZZZ$ measurements without direct sensitivity to unspecified UV cut-offs, whose role is taken over by the physical mass scales of a concrete UV simplified theory. The merit of simplified models is hence two-fold: Firstly, they provide a minimal interface that captures both resonant and non-resonant features of an BSM-motivated scenario. Secondly, they provide a framework to critically assess the sensitivity reach of colliders, allowing a direct comparison of different collider concepts within a consistent theoretical framework. The price paid for this level of predictability is that we are limited to perturbative theories, which possess a well-defined approach to renormalisation and power-counting of interactions.

This work is organised as follows: we first discuss the anomalous coupling parameterisation of a tree-level $ZZZ$ vertex and consider the size of the one-loop induced effect in the SM. We then consider the one-loop induced effect from additional scalars and fermions, using a 2HDM and a fourth generation of vectorlike leptons\footnote{We focus on vectorlike leptons as constraints on coloured particles such as top partners~\cite{Matsedonskyi:2015dns} are substantially stronger than for vectorlike leptons~\cite{Kumar:2015tna}} as example simplified models to probe the sensitivity of $ZZZ$ measurements in minimal BSM scenarios beyond the region of applicability of EFTs. We find that the prospects for such measurements at hadron colliders are small, given the large SM backgrounds. While measurable deviations are possible at lepton colliders, we find that these are due to the effects of the new particles on polarisation functions rather than the $ZZZ$ vertex, making it difficult to tie deviations in $e^+e^- \to ZZ$ to new sources of $CP$ violation.

\section{Anatomy of $ZZZ$ interactions}
\label{sec:anatomy}
In the SM there are no tree level $ZZZ$ couplings but these interactions are generated from UV-finite one-loop corrections.
Accordingly, we consider scenarios in which the $ZZZ$ couplings are generated radiatively and focus on scalar and fermionic degrees of freedom, taking inspiration from the SM. Beginning from Lorentz invariance and requiring two $Z$ bosons be on shell, the general form of the triple--$Z$ coupling may be written as \cite{Gounaris:2000tb}:
\begin{multline}
\label{eq:definef45}
\Gamma^{\alpha\beta\mu}(q_1,q_2,P)=\frac{i(P^2-m_Z^2)}{m_Z^2}\left[f_4^Z(P^\alpha g^{\mu\beta}+P^\beta g^{\mu\alpha})\right.\\
  \phantom{\frac{i(P^2-m_Z^2)}{m_Z^2}}\left.-f_5^Z\epsilon^{\mu\alpha\beta\rho}(q_1-q_2)_\rho\right]\, .
\end{multline}
where $P^\mu$ is the incoming off--shell momentum, $q_1^\alpha$ and $q_2^\beta$ are the outgoing on--shell momenta. The form factor $f_4^Z$ is $CP$--violating while $f_5^Z$ preserves $CP$. In general the form factors $f_{4,5}^Z$ are nonanalytic functions of the momenta which contain information about the loop dynamics that generate the $ZZZ$ interaction. The behavior of these form factors as a function of center of mass energy has been studied in the SM and a variety of BSM models from a phenomenological perspective in Refs.~\cite{Gounaris:2000tb,Chang:1994cs,Baur:2000ae,Gounaris:1999kf,Grzadkowski:2016lpv,Grzadkowski:2014ada}.

These form factors have also been studied by the ATLAS and CMS Collaborations~\cite{Aaboud:2016urj,CMS:2014xja,Khachatryan:2015pba,CMS:2017ruh,Aaboud:2017rwm,Sirunyan:2017zjc}, who place bounds on the size of the form factors in Eq.~\eqref{eq:definef45}. In these analyses the LHC experiments typically treat the form factors as constants and provide limits on their size, neglecting the dependence on the involved momentum scales which is in general model dependent. The omission of any momentum dependence in the form-factors beyond Lorentz-symmetry considerations leads to unitarity violation, which is often tamed through the introduction of form factors (see e.g.~\cite{Hagiwara:1986vm,Baur:1987mt,Baur:1988qt,Baur:2000ae})
\begin{equation}
f_{4,5}^{Z}\to f_{4,5}^Z\left(1+\frac{P^2}{\Lambda^2}\right)^{-2}\, .
\end{equation}
These choices serve to fully ameliorate the effects of the anomalous triple--$Z$ coupling at high $P^2$.
This technique introduces further model dependence into the interpretation of $f_{4,5}^Z$ as not all models are unitarised identically.
If nothing is done to quell unitarity violation, overly stringent constraints will be obtained as limit-setting will be driven by the unsuppressed signal cross sections for large $p_T$ bins. In this sense it is hard to gauge whether the limit is a result of perturbative unitarity or really relates to the lack of new physics, which could be well-described by perturbative means. The most stringent constraints are derived in this manner, namely,~Ref.~\cite{Sirunyan:2017zjc},
\begin{equation}\label{eq:cmsf45z}
\renewcommand\arraystretch{1.5}
\begin{array}{rcccl}
-0.0012&<&f_4^Z&<&0.0010\, ,\\
-0.0010&<&f_5^Z&<&0.0013\, .
\end{array}
\renewcommand\arraystretch{0}
\end{equation}
Comparison of these results with similar limits obtained at LEP, e.g. recent L3 results \cite{Achard:2003hg},
\begin{equation}\label{eq:lepf45z}
\renewcommand\arraystretch{1.5}
\begin{array}{rcccl}
-0.48&<&f_4^Z&<&0.46\, ,\\
-0.36&<&f_5^Z&<&1.03\, .
\end{array}
\renewcommand\arraystretch{0}
\end{equation}
D0~\cite{Abazov:2007ad}
\begin{equation}\label{eq:lepf45zd0}
\renewcommand\arraystretch{1.5}
\begin{array}{rcccl}
-0.28&<&f_4^Z&<&0.28~(\Lambda=1.2~\text{TeV})\, ,\\
-0.31&<&f_5^Z&<&0.29~(\Lambda=1.2~\text{TeV})\, .
\end{array}
\renewcommand\arraystretch{0}
\end{equation}
or ATLAS~\cite{Abazov:2007ad}
\begin{equation}\label{eq:lepf45zatl}
\renewcommand\arraystretch{1.5}
\begin{array}{rcccl}
-0.019&<&f_4^Z&<&0.019~(\Lambda=3~\text{TeV})\, ,\\
-0.020&<&f_5^Z&<&0.019~(\Lambda=3~\text{TeV})\, .
\end{array}
\renewcommand\arraystretch{0}
\end{equation}
indicates how much the kinematic coverage feeds into the constraints once potential energy-dependencies are not considered. Identifying $f_4^Z$ and $f_5^Z$ with effective operators we can cast these constraints into new physics scales in the context of an effective field theory.

Beginning with $f_4^Z$ we follow~\cite{Degrande:2013kka} which finds that three different dimension--eight operators contribute to $f_4^Z$ (there is no contribution at dimension--six):
\begin{equation}
f_4^Z=\frac{M_Z^2v^2}{2c_Ws_W}\frac{(c_W^2c_{WW}+2c_Ws_Wc_{BW}+4s_W^2c_{BB})}{\Lambda^4}\, ,
\end{equation}
where $s_W^2\equiv1-c_W^2\equiv\sin^2\theta_W$ is the Weinberg angle, and the Wilson coefficients $c_{WW}$, $c_{BW}$, and $c_{BB}$ correspond to the effective operators,
\begin{align}\label{eq:d8ZZZops}
\mathcal{O}_{WW}&=iH^\dagger W_{\mu\nu}W^{\mu\rho}\{D_\rho,D^\nu\}H\, ,\\
\mathcal{O}_{BW}&=iH^\dagger B_{\mu\nu}W^{\mu\rho}\{D_\rho,D^\nu\}H\, ,\\
\mathcal{O}_{BB}&=iH^\dagger B_{\mu\nu}B^{\mu\rho}\{D_\rho,D^\nu\}H\, .
\end{align}
If we assume only one of the operators is generated by a new UV complete model with a Wilson coefficient $c_i\sim1$ we can infer the scale of new physics from the maximum allowed size of $f_4^Z$ given above. We find the lowest scale of new physics (NP) corresponds to the operator $\mathcal{O}_{BB}$ giving a scale $\Lambda\sim680$ GeV. Since this is a loop generated effect, if we take instead a loop-suppressed Wilson coefficient $c_i\sim1/(16\pi^2)$ we find a lowest scale of $\Lambda_{\rm NP}\sim190$ GeV.
Next we can connect $f_4^Z$ with a dimension twelve operator, which was recently identified in~\cite{Belusca-Maito:2017iob},
\begin{multline}
\mathcal{O}_{4Z}=\frac{c_{4Z}}{\Lambda^8}(H^\dagger D_\mu H)^2(H^\dagger D_\nu H)^2+{\text{h.c.}}\, .
\end{multline}
This operator will generate a $Z^3 \partial h$ vertex at tree level when expanded, which allows
a $ZZZ$ contribution to be induced at one loop. Therefore we take $c_{4Z}\sim1$, assuming a loop suppression factor in the IR theory, and using the bounds in Eq.~\eqref{eq:cmsf45z} we find a NP scale corresponding to $\Lambda_{\rm NP}\sim200$ GeV. It is important to note that generic UV complete models typically generate more than one effective operator so these derived scales should be taken as a guide only \cite{Jiang:2016czg,Corbett:2017ieo}.
$f_5^Z$ is not generated at dimension-six in SM EFT framework~\cite{Dedes:2017zog} (however, there are are similar $WWZ$ interactions~\cite{Hagiwara:1986vm}).  Therefore constraints from $f_5^Z$ are likely to impose constraints on BSM scenarios which are comparable to those from the above discussions of $f_4^Z$.

Given this we see that despite the seemingly strong constraints the experiments have placed on $f_{4,5}^Z$ they do not indicate strong constraints on the mass scale of new physics. Given the relatively low constraints it is possible that the new degrees of freedom are propagating and an EFT or constant form factor approach is not appropriate. We thus adopt a more UV-complete perspective testing extended scalar and fermion sectors, which also covers potentially large threshold effects.

In the following we will consider the full one--loop expressions\footnote{That is, we do not use the expressions in Eq.~\eqref{eq:definef45}, but instead derive the full one loop dependence, not assuming any legs are on-shell. We use this full form of the vertex for all LHC simulations which follow, including those with on shell final state $Z$s.} for the $ZZZ$ vertex in the SM and different New Physics scenarios, and discuss the expected size of the contribution of the $ZZZ$ vertex to the $pp\to ZZ$ and $pp\to Zjj$ processes at the $13$ TeV LHC and future linear colliders. We will specifically focus on the potential effects of thresholds that might provide a sensitive probe of new physics. It should be stressed however, that the LHC cross sections provided in this work should be understood as approximations to the full electroweak corrections in these modified scenarios in the sense that we add finite contributions which are loop-induced and have no leading-order counterpart, i.e. they are new partonic subprocesses \footnote{Similar strategies have been applied for associated Higgs production from gluon fusion,~see e.g.~Refs.~\cite{Kniehl:2011aa,Altenkamp:2012sx,Harlander:2013mla,Englert:2013vua}}. For the case of the LHC, full NLO electroweak corrections have been provided recently in~Refs.\cite{Bierweiler:2013dja,Baglio:2013toa,Denner:2014ina,Biedermann:2016lvg,Biedermann:2016yvs}.

Since these processes are loop-induced and electroweak in nature, their effects can be small at the LHC; for the more promising case of the lepton models we will therefore also discuss the expectations at a future lepton collider where increased sensitivity will open the possibility to probe such new states through their modified electroweak corrections.

\subsection{Within the Standard Model}
\label{sec:anatomysm}

We first consider the generation of the $ZZZ$ vertex at one-loop within the SM. Most loop-contributions cancel exactly and the only contributions which do not vanish at one--loop are those from intermediate fermions in the loop~\cite{Gounaris:2000tb,Gounaris:2000dn,Gounaris:1999kf}. We can understand this behavior by considering the SM current which couples the $Z$--boson to the other fields of the SM.

We are therefore left only with the possibility of a contribution due to the fermions. Due to their properties under $SU(2)_L$ the $t$-- and $b$--quark loops we find they destructively interfere to give a suppressed overall cross section \cite{Gounaris:2000dn}. Employing {\sc FeynRules}~\cite{Christensen:2008py,Alloul:2013bka} (with output in the {\sc Ufo} format~\cite{Degrande:2011ua}), {\sc FormCalc}~\cite{Hahn:1998yk,Hahn:2000kx}, and {\sc{MadGraph5}}~\cite{Alwall:2011uj,Alwall:2014hca} we generate cross sections for the SM $ZZ$ production process.

As expected \cite{Gounaris:2000dn}, the one--loop $ZZZ$ coupling is at most a $0.05\%$ deviation for the $ZZ$ process which should be contrasted with a theoretical uncertainty of 3\% on the Standard Model cross section at NNLO QCD \cite{Heinrich:2017bvg}, and an experimental uncertainty of about 5\% in the latest ATLAS and CMS measurements \cite{Aaboud:2017rwm,Sirunyan:2017zjc}. While large improvements in both uncertainties can be expected by the end of the HL-LHC physics programme, these are unlikely to be sufficient to make the one--loop $ZZZ$ vertex in the Standard Model measurable in this channel.

Alternatively, one can consider the the weak boson fusion component of $pp\to Zjj$ production that is selected through weak boson fusion cuts~\cite{Rainwater:1996ud}\footnote{We adopt jet transverse momentum cuts of $p_{T,\text{j}}>20~\text{GeV}$, azimuthal-angle--pseudorapidity separation $\Delta R_{\rm jj}>0.4$, a large rapidity separation of the jets $|\delta y_{\rm{jj}}|>4$ at large jet-invariant transverse mass $m_{\text{jj}}>400~\text{GeV}$. For these criteria the $Z$ boson decays centrally, with no additional efficiency suppression from lepton-isolation criteria for $Z\to \ell^+\ell^-$. We report cross section numbers that do not include the $Z$ decay branching ratios throughout this paper, and indicate the weak boson fusion contribution with a 'WBF' superscript.}, where the interactions under discussion do introduce new partonic subprocesses compared to the born level calculation. These contributions, however, are suppressed by a factor of $10^6$ relative to the leading ones.

Therefore within the SM it is unlikely any progress can be made on measuring the one--loop $ZZZ$ process at the LHC. As differential electroweak corrections are of the order of 10\%~\cite{Denner:2014ina,Biedermann:2016yvs,Biedermann:2016lvg}, typically with a non-trivial interplay with QCD contributions in the case of $Zjj$ production~\cite{Denner:2014ina}, lepton colliders are a particularly motivated environment to test the presence of new electroweak states indirectly.

\subsection{Scalars and the $CP$--violating 2HDM}
\label{sec:2HDM}
We consider the addition of new scalars and fermions which couple to the $Z$-boson. New vector resonances may also contribute, but since we focus on perturbative completions in this work, we will not focus on them any further. We note that some discussion of vector resonances can be found in~\cite{Gounaris:2000tb}. In this subsection we consider the affect of extended scalar sectors, focusing on the 2HDM as a particular example, and move on to additional fermions in Sec.~\ref{sec:vll}.

The simplest extended scalar sectors involve the addition of one new $N$--plet of $SU(2)_L$ with some hypercharge $Y$. The $U(1)_Q$ charge of a particular component of the new scalar is given by
\begin{equation}
Q=T^3+Y\, ,
\end{equation}
where $T^3$ is the diagonal generator of $SU(2)_L$ in the $N$--dimensional representation. There will be only one neutral component of the $N$--plet for a given hypercharge. For example, for a real scalar in the $N$ representation of $SU(2)_L$ we expect $(N-1)/2$ charged scalars and one $CP$--even neutral scalar\footnote{One may only form a real scalar of an odd dimensional representation of $SU(2)_L$. For a real scalar the hypercharge is necessarily 0, and therefore the $(N-1)/2+1$th component of the scalar is neutral and all others are charged. The remaining components are $(N-1)$ charged scalars of which $(N-1)/2$ are necessarily related to the others by charge conjugation.}. For any complex scalar we expect either $(N-1)$ charged scalars and one $CP$--even and one $CP$--odd neutral scalar, or $N$ charged scalars, depending on the hypercharge assignment.

The additional charged scalars will be subject to Furry's theorem and will not contribute to the $ZZZ$ loop. One might expect contributions of the neutral components will vanish similar to the case of the SM Higgs. However, since there are additional neutral scalars in the Lagrangian, mixing effects allow this issue to be evaded. We will consider the 2HDM as an example of extended scalar sectors. Since, as mentioned above, all extended scalar sectors with a single new $N$--plet of $SU(2)_L$ will have at most one additional $CP$--even and one $CP$--odd neutral scalar, we take the 2HDM and its phenomenology to be representative of all models in this class.

Our discussion of the 2HDM will follow the work \cite{Grzadkowski:2014ada,Grzadkowski:2016lpv} which discusses the $ZZZ$ vertex resulting from the $CP$--violating 2HDM. The 2HDM scalar potential is
\begin{multline}
V=-\frac{m_{11}^2}{2}|\Phi_1|^2-\frac{m_{22}^2}{2}|\Phi_2|^2-\frac{1}{2}\left(m_{12}^2\Phi_1^\dagger\Phi_2+{\text{h.c.}}\right)\\
+\frac{\lambda_1}{2}|\Phi_1|^4+\frac{\lambda_2}{2}|\Phi_2|^4+\lambda_3|\Phi_1|^2|\Phi_2|^2\\
+\lambda_4(\Phi_1^\dagger\Phi_2)(\Phi_2^\dagger\Phi_1)+\frac{1}{2}\left(\lambda_5(\Phi_1^\dagger\Phi_2)^2+{\text{h.c.}}\right)\\
+\left(\left[\lambda_6(\Phi_1^\dagger\Phi_1)+\lambda_7(\Phi_2^\dagger\Phi_2)\right](\Phi_1^\dagger\Phi_2)+{\text{h.c.}}\right)\, .
\end{multline}
Possible complex parameters of this Lagrangian include
\begin{equation}
\left\{m_{12},\lambda_5,\lambda_6,\lambda_7,e^{i\xi}\right\}\, ,
\end{equation}
where $\xi$ is the relative phase between the vevs of $\Phi_1$ and $\Phi_2$. Of these complex parameters an overall $SU(2)$ rephasing of the scalar potential may remove two phases leaving a total of three complex parameters in the model~\cite{Gunion:2005ja}.

Expanding  $\Phi_1$ and $\Phi_2$ in terms of their component fields and vacuum expectation values $v_i$,
\begin{equation}
\Phi_i(x)=e^{i\xi_i}\left(\renewcommand\arraystretch{1.5}\begin{array}{c}\phi_i^+(x)\\(v_i+\eta_i(x)+i\chi_i(x))/\sqrt{2}\end{array}\renewcommand\arraystretch{0}\right)
\end{equation}
results in mixing between the various components. In order to obtain the physical states we first rotate to a basis with massless Goldstone boson fields $G^0$ and $G^{\pm}$, and massive physical states $\eta_3$ and $H^\pm$ via
\begin{equation}
\renewcommand\arraystretch{1.3}
\left(\begin{array}{c}G^0\\ \eta_3\end{array}\right)=\left(\begin{array}{cc}v_1/v&v_2/v\\-v_2/v&v_1/v\end{array}\right)\left(\begin{array}{c}\chi_1\\ \chi_2\end{array}\right)\, ,
\renewcommand\arraystretch{0}
\end{equation}
and
\begin{equation}
\renewcommand\arraystretch{1.3}
\left(\begin{array}{c}G^{\pm}\\ H^{\pm}\end{array}\right)=\left(\begin{array}{cc}v_1/v&v_2/v\\-v_2/v&v_1/v\end{array}\right)\left(\begin{array}{c}\phi_1^{\pm}\\ \phi_2^{\pm}\end{array}\right)\, .
\renewcommand\arraystretch{0}
\end{equation}
If $CP$ is conserved there is  mixing between the two $CP$--even scalars $\eta_1$ and $\eta_2$, but not with the $CP$-odd $\eta_3$ state. The theory has couplings $Z\eta_1 \eta_3$ and $Z\eta_2\eta_3$ between the $Z$ and a $CP$--even and the $CP$--odd scalar, as well as $ZZ\eta_1$ and $ZZ\eta_2$ couplings. However these are still insufficient to generate the $ZZZ$ vertex at one loop as there is no $Z\eta_1\eta_2$ or $Z\eta_i\eta_i$ coupling. For any real $N$--plet, or complex $N$--plet without $CP$--violation there is no contribution to the $ZZZ$ vertex at one--loop.

\begin{figure}
\includegraphics[width=6.6cm]{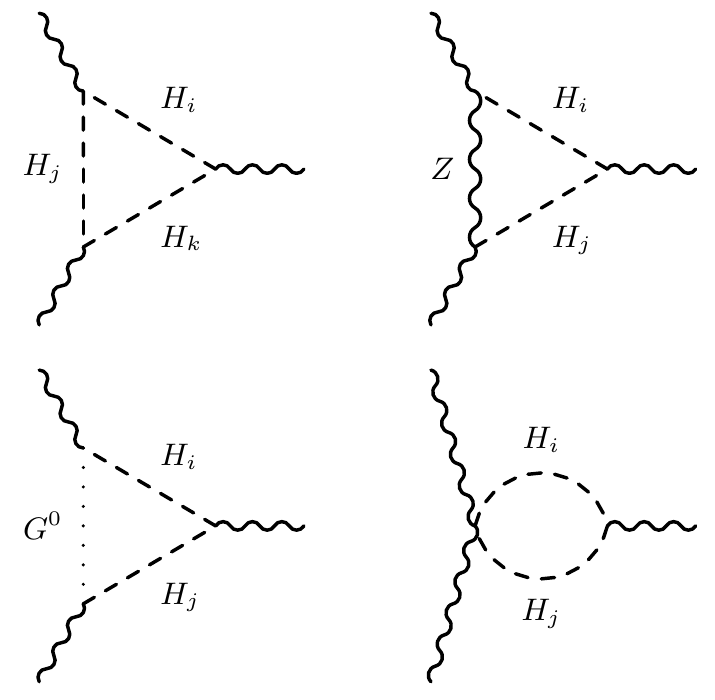}
\caption{The four graphs involving scalars contributing to the $ZZZ$ vertex in the Two Higgs Doublet Model. Additional contributions come from the SM fermions as discussed in Sec.~\ref{sec:anatomysm}. In all diagrams $i$, $j$, and $k$ must be different. The bubble diagram (bottom right) is identically zero, but included for completeness. We agree with the relative sign of the diagram involving the $Z$ in the loop that was recently corrected in \cite{Belusca-Maito:2017iob}.}\label{fig:ZZZ2HDM}
\end{figure}

However, for a $CP$--violating 2HDM there will generally be mixing between the three neutral components $\eta_i$. The mass matrix for the neutral states is diagonalised by an orthogonal mixing matrix $R$,
\begin{equation}
\left(\begin{array}{c}H_1\\ H_2\\ H_3\end{array}\right)=R\left(\begin{array}{c}\eta_1\\ \eta_2\\ \eta_3\end{array}\right)\, ,
\end{equation}
where the $H_i$ are the physically propagating states. After this rotation there are three scalars of mixed $CP$. The Lagrangian coupling the scalars to the $Z$--boson now has the couplings $ZH_iH_j$ (for $i\ne j$), $ZH_iG^0$, and $Z^2H_i$. There are also quartic interactions between two scalars and two gauge bosons, however diagrams involving this coupling are identically zero.
Using these interactions one can construct all of the diagrams in Fig.~\ref{fig:ZZZ2HDM} which contribute to the $ZZZ$ vertex at one loop in the 2HDM (in addition to the SM contributions previously discussed). We note that our calculation agrees with the results recently obtained by~\cite{Belusca-Maito:2017iob}.

With this framework in hand we are now free to proceed to calculate the loops and simulate their effects as in Sec.~\ref{sec:anatomysm}. We begin by noting that all diagrams (except the bubble which is identically zero) in Fig.~\ref{fig:ZZZ2HDM} share the common prefactor~\cite{Grzadkowski:2014ada,Grzadkowski:2016lpv}:
\begin{multline}
c_{ZZZ}\equiv\\\frac{(R_{11}v_1+R_{12}v_2)(R_{21}v_1+R_{22}v_2)(R_{31}v_1+R_{32}v_2)}{v^3}\,.
\end{multline}
We consider $c_{ZZZ}$ as a function of the mixing angles $\theta_i$ in the matrix $R$ and the vevs $v_i$, and maximise its value subject to $v_1^2+v_2^2=(246\ \mbox{GeV})^2$ and $-\pi/2 \leq \theta_i \leq \pi/2$. We find that $c^{max}_{ZZZ}\sim 0.19$
We assume this maximal value and simulate the size of the one--loop contribution to both the $ZZ$ and $Zjj$ processes. As the effects are too small to distinguish from the SM contribution we only give here the size of the loop contribution squared to demonstrate how small the effect is. Scanning over $M_{H_2}\in (250,1250)$ GeV and $M_{H_2}<M_{H_3}\in(500,1500)$ GeV the maximal values of the cross section for both processes are
$\Delta \sigma_{ZZ}\sim\mathcal{O}(10^{-3})~{\rm fb}$ and $\Delta \sigma_{Zjj}^{\text{WBF}}\sim\mathcal{O}(10^{-3}) {\rm fb}$.
These values are too small to be probed at the LHC in particular as the associated QCD uncertainty is at the level of 3\% for $Z$ pair production total cross section of $\sim 17~\text{pb}$ \cite{Heinrich:2017bvg} or a WBF $Zjj$ cross section of $\sim 4~\text{pb}$~\cite{Oleari:2003tc,Jager:2012xk,Re:2012zi,Schissler:2013nga,Denner:2014ina}. Percent-level corrections can be expected from non-resonant modified electroweak corrections (see below) but these have multiple sources and do not pinpoint a $ZZZ$ vertex. Therefore, the 2HDM is not the optimal UV complete motivation for studying the loop induced $ZZZ$ vertex. Indeed inspecting~\cite{Grzadkowski:2016lpv}, while cross sections are not provided but elements of the amplitude are discussed, we would expect similar cross sections since~\cite{Grzadkowski:2016lpv} shows
\begin{equation}
f_4^Z\sim 10^{-4}\,\delta\, ,
\end{equation}
where $\delta=1-(v_1R_{11}+v_2R_{12})/v\lesssim 5\%$~\cite{Grzadkowski:2014ada}. This is significantly below the constraint from CMS in Eq.~\eqref{eq:cmsf45z}, a constraint which itself may be relaxed by a more realistic treatment of the high $p_T$ part of phase-space as mentioned in the beginning of this section.

As we have previously argued any scalar sector extended by at most one complex $N$--plet of $SU(2)_L$ only allows for at most two new neutral scalars. In any such scenario the new scalar will contribute a similarly negligible contribution to the $ZZZ$ coupling. That is to say, any extension of the SM scalar sector by a complex $N$--plet will generate a negligible contribution to the $ZZZ$ coupling at one loop and therefore may be disregarded phenomenologically for studies of the $ZZZ$ coupling.

\subsection{Simplified Fermionic Models}
\label{sec:vll}
In the light of the previous discussion, the first fermionic simplified model that one could consider is a single fermion with axial $U(1)_Y$ couplings
\begin{equation}
{\cal{L}}_\psi = i \bar{\psi} \gamma^\mu(\partial_\mu - i\alpha g'B_\mu  \gamma_5)\psi - m_\psi \bar \psi \psi
\label{eq:singlet}
\end{equation}
which leads to a QED charge for the field $\psi$ of $\alpha$ after rotating the hypercharge gauge field $B$ to the gauge boson mass basis. Such a state decouples leading to vanishing oblique corrections. However, the one-loop $AZZ$ and $ZZZ$ interactions which are relevant for $ZZ$ production as well as VBF $Zjj$ production also vanish and such a model does not lead to an interesting new physics signal for our purposes.

The only way to include sensitivity to thresholds while keeping the possibility to compare to oblique electroweak corrections is by introducing additional ``chiral'' masses through the Higgs mechanism on top of vectorlike masses. The effects discussed in the context of the third SM family of quarks can then be lifted to a higher mass scale and comparably large non-diagonal $Z$ couplings of the fermions in the mass basis can be induced in principle. We take this as motivation to consider a fourth generation of vectorlike leptons as another minimal and concrete BSM scenario with potential sensitivity to $ZZZ$ measurements. Such scenarios have been discussed in the context of $H\to \gamma \gamma$ measurements~\cite{Joglekar:2012vc} and they provide an avenue to raise the mass of the lightest Higgs boson in models of weak-scale supersymmetry, since the mass correction from new vectorlike supermultiplets will be positive if the fermions are lighter than their scalar partners \cite{Moroi:1991mg,Moroi:1992zk,Babu:2004xg}. The mass spectrum is determined by the vectorlike mass terms and Yukawa couplings given by
\begin{align}
  \label{eq:mass4l}
 - \mathcal{L}_{\text{mass}} \supset m_{l} \bar{l}_{L}' l_{R}'' + m_{e} \bar{e}_{L}'' e_{R}' + m_{\nu} \bar{\nu}_{L}'' \nu_{R}' + &\text{ h.c.} \nonumber \\
    + Y_{c}'(\bar{l}_{L}' H) e_{R}' + Y_{c}''(\bar{l}_{R}'' H) e_{L}'' + &\text{ h.c.} \nonumber \\ + Y_{\nu}'(\bar{l}_{L}' \widetilde{H}) \nu_{R}' + Y_{\nu}''(\bar{l}_{R}'' \widetilde{H}) \nu_{L}'' + &\text{ h.c.}\,.
\end{align}
Here $\widetilde{H} = i \sigma^2 H^\dagger$ and all coupling parameters are chosen to be real. All of the fields are singlets under $SU(3)_C$ and their $SU(2)_L \times U(1)_Y$ charges are given in Tab.~\ref{tab:vlikecharges}. Unlike a new fermion generation with only Yukawa coupling-induced mass terms, the electroweak singlet mass terms allow the vectorlike fermions to decouple from electroweak precision constraints and on-shell Higgs observables~\cite{Joglekar:2012vc}.

\begin{table}
\begin{tabular}{ l | c | c | c}
Field & $l_{L}'$, $l_{R}''$ & $e_{L}''$, $e_{R}'$ & $\nu_{L}''$, $\nu_{R}'$\\
\hline
$SU(2)_L \times U(1)_Y$ & $(\textbf{2},-1/2)$ & $(\textbf{1},-1)$ & $(\textbf{1},0)$\\
\end{tabular}
\caption{The quantum numbers of the fields of the new lepton generation under $SU(2)_L \times U(1)_Y$.}\label{tab:vlikecharges}
\end{table}

After electroweak symmetry breaking the Lagrangian leads to $2 \times 2$ mixing matrices in the charged and neutral sectors:
\begin{multline}
 - \mathcal{L}_{\text{mass}} \supset \begin{pmatrix} \bar{e}_{L}' & \bar{e}_{L}'' \end{pmatrix}
    \begin{pmatrix} \frac{v Y_{c}'}{\sqrt{2}} & m_{l} \\ m_{e} & \frac{v Y_{c}''}{\sqrt{2}} \end{pmatrix}
    \begin{pmatrix} e_{R}' \\ e_{R}'' \end{pmatrix} \\
    + \begin{pmatrix} \bar{\nu}_{L}' & \bar{\nu}_{L}'' \end{pmatrix}
    \begin{pmatrix} \frac{v Y_{\nu}'}{\sqrt{2}} & m_{l} \\ m_{\nu} & \frac{v Y_{\nu}''}{\sqrt{2}} \end{pmatrix}
    \begin{pmatrix} \nu_{R}' \\ \nu_{R}'' \end{pmatrix}
\end{multline}

\begin{figure}
\includegraphics[width=8cm]{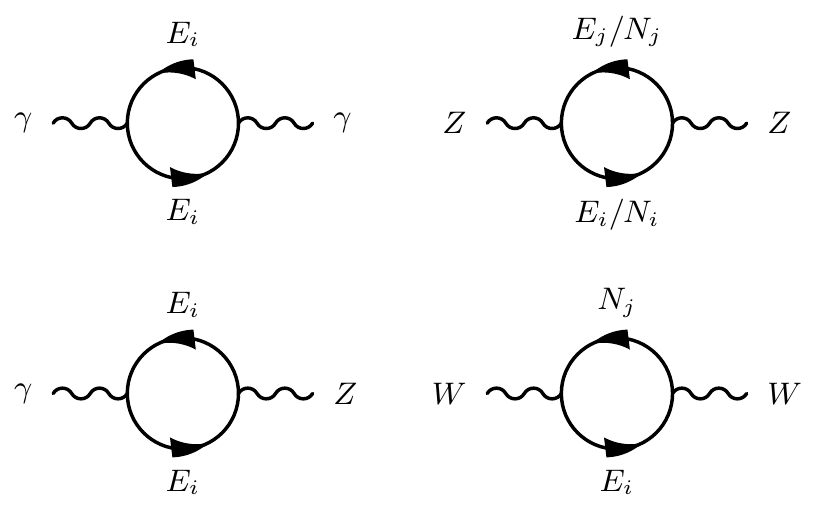}
\caption{The polarisation diagrams which contribute to the $S, T, U$ parameters \cite{Peskin:1991sw} in the vectorlike lepton model.  }\label{fig:STU}
\end{figure}
%
\begin{figure*}[!t]
	\includegraphics[width=0.45\textwidth]{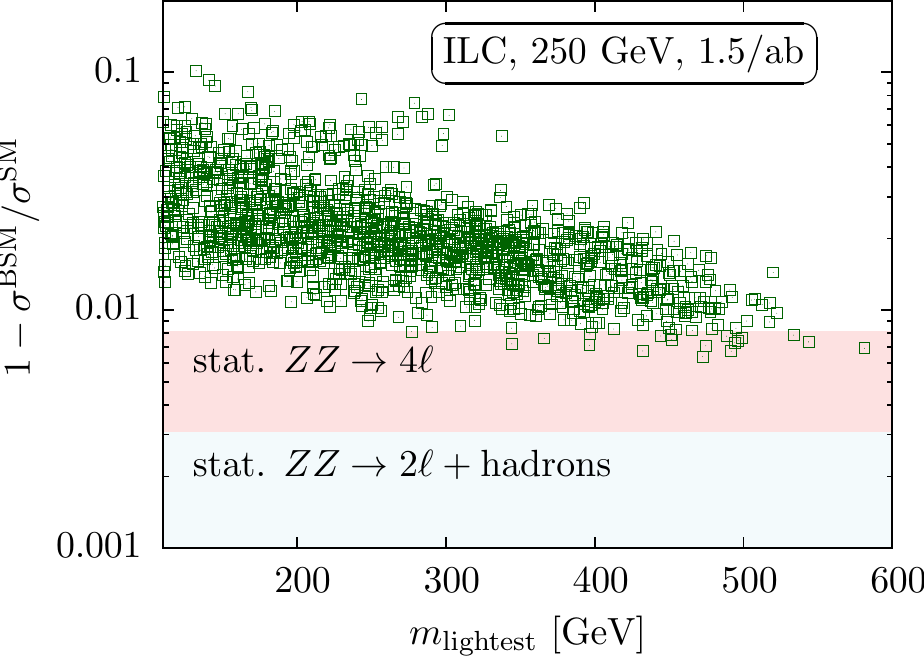}\hskip0.8cm
	\includegraphics[width=0.45\textwidth]{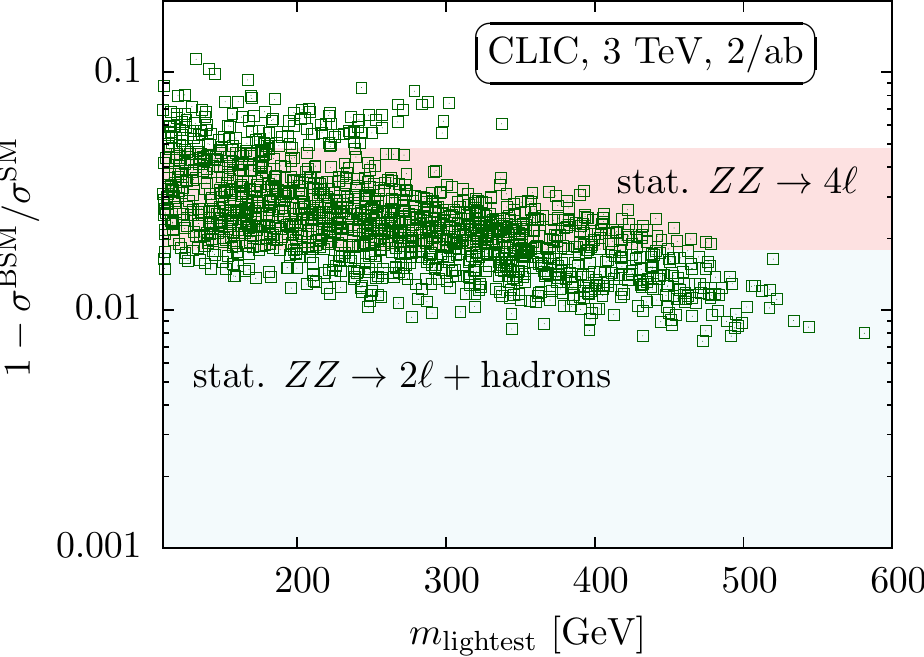}
	\caption{\label{fig:leptoncollider}Relative size  of the vectorlike lepton contribution to the $ZZ$ cross section with respect to the Standard Model expectation for centre-of-mass energies corresponding to the ILC and CLIC colliders. The results are based on full on-shell electroweak NLO calculations in the on-shell renormalisation scheme as implemented in \cite{Hahn:1998yk,Hahn:2000kx}. The parameter points are randomly distributed over the parameter space, and demonstrate that the cross section can vary by an order of magnitude for a fixed value of the lightest fermion mass eigenstate $m_\text{lightest}$. All points pass the $S,T,U$ constraints of Ref.~\cite{Baak:2012kk} at the 95\% confidence level. The red and blue bands are based on statistical uncertainties of $ZZ$ measurements assuming the Standard Model using the leptonic and semi-leptonic final states and expected end-of-lifetime luminosities for the machines. We assume on-shell $Z$ production throughout; no systematic uncertainties are included in these figures. 
	}
\end{figure*}
%
Rotating from the Lagrangian eigenstates $( e_{L}', e_{L}'', e_{R}', e_{R}'')$ and $( \nu_{L}', \nu_{L}'', \nu_{R}', \nu_{R}'')$ to the mass eigenstates $(E_1, E_2)$ and $(N_1,N_2)$ will determine the relevant gauge interactions.
In order to observe $C$ violation none of the parameters is allowed to vanish, i.e. in our scan we need to vary all seven parameters independently. Zero entries (or simplifications through identifications of parameters) lead to a vanishing $ZZZ$ interaction. In such a case the presence of the fermions would still manifest itself in precision observables as well as in non-oblique corrections of collider cross sections. In fact these can overpower the phenomenological cross section-suppression detailed in~\cite{Belusca-Maito:2017iob}, however, at the price of a loss of direct interpretation in terms of $C$ or $CP$ violation, see below.

In the spirit of using simplified models to cross-relate different measurements some comments are in order. Since we do not mix the new lepton generation with the Standard Model leptons, there is in principle a parity symmetry protecting decays to the Standard Model which will make the lightest mass eigenstate stable on cosmological timescales and it will contribute a relic density. A charged exotic relic density should be avoided, so taking this effect at face value, we would have to require the lightest mass eigenstate to be $N_{1/2}$. A dark matter interpretation would however additionally have to avoid overclosing the universe while escaping direct detection constraints. These have been studied, with the addition of Majorana mass terms which split the $N_i$ into four mass eigenstates, in \cite{Joglekar:2012vc} which found viable parts of parameter space, relying on the Xenon100 results~\cite{Aprile:2012nq} as their most-constraining spin-independent limits. Xenon1T has recently improved these constraints by an order of magnitude~\cite{Aprile:2017iyp} compared to those used in this previous study, which forces the model to rely on co-annihilation between the lightest $N_i$ and $E_i$ and hence additionally requires one of the charged scalars to be close to mass degenerate with the $N_i$ which forms the relic density.

Since we do not want to constrain our parameter space to this extent and rather study it in a more general manner, while avoiding flavor-changing interactions involving $e$ and $\mu$, we will assume there is a small mixing with the third lepton generation in the Standard Model and the new vectorlike generation which can be ignored for the purpose of our calculations. This will avoid dark matter constraints completely and make decays of pair produced $E_i/N_i \to \tau/\nu_\tau + h/Z/W$ interesting direct signatures which can be looked for at colliders. Since LEP failed to find any such signatures this puts a lower bound on the lowest mass eigenstate $m_{\text{lightest}} > 100.8$ GeV (102.6 GeV in case there is no mixing with the third lepton generation) when the lightest state is charged and $m_{\text{lightest}} > 90.3$ GeV when the lightest state is neutral (this weaker limit is the result of only mixing with the third lepton generation, and would rise to 101.5 GeV if the decay to $\mu W$ was favoured instead) \cite{Achard:2001qw}. These can only can be avoided if the mass mixing suppresses the coupling of the lightest state to the $Z$.

Higher masses are sensitive to direct searches at the LHC. These were studied in a phenomenological context in \cite{Kumar:2015tna}. However the two benchmarks models considered in their study correspond roughly to the $SU(2)_L$ singlet (and doublet models) of Eqs.~\eqref{eq:singlet} which have a simpler mass spectrum and interaction structure than the model we consider, and we can not easily recast their limits. To do so would require a propagation of mixing effects to both the production cross section and branching ratio calculations, taking into account new decays such as $E_1 \to N_1 W$ which are absent in the mass degenerate case. Such a study is outside the scope of this paper, but based on the previous work we can reasonably expect the LHC to be sensitive to $m_{\text{lightest}}$ in the range of several hundred GeV. In the limit of sequential couplings with the lightest charged and neutral states mass degenerate we can apply the results from \cite{Kumar:2015tna} which imply a potential discovery range up to $m_{\text{lightest}} = 450$ GeV with the full HL-LHC dataset in this corner of the full parameter space.

More concrete constraints that relate to the generic modification of the electroweak sector due to the new states can be imposed through oblique corrections that arise from the model of Tab.~\ref{tab:vlikecharges} and Eq.~\eqref{eq:mass4l}. The diagrams contributing to the $S,T,U$ parameters via the weak gauge boson polarisation functions are given in Fig.~\ref{fig:STU} and the resulting constraints on the model's parameter space have been studied in Ref.~\cite{Joglekar:2012vc,Englert:2013tya}. We scan the model over the relevant parameters in  Eq.~\eqref{eq:mass4l} and keep parameter points that are in agreement with the constraints of \cite{Baak:2012kk} at the 95\% confidence level. In the following we will project these results onto the mass of the lightest state of Eq.~\eqref{eq:mass4l} after diagonalisation.

The impact of the new fermion loop contributions to the $ZZZ$ vertex on $ZZ$ and $Zjj$ production at the LHC, although bigger than in the two Higgs doublet model, is again small upon comparison with the SM. To illustrate this, we consider a parameter point whose \textit{total} electroweak corrections (i.e.~including both SM and BSM fields) are $\sim 15\%$ at a lepton collider (to be discussed below) with a small dependence on energy, corresponding to a lightest vector-like lepton mass of $\sim~300~\text{GeV}$. In this case, if we take into account only the effects of the $ZZZ$ triangle diagrams we find changes in the cross-sections of
\begin{equation}
\renewcommand\arraystretch{1.5}
\begin{array}{rcrl}
\Delta \sigma_{ZZ}&\lesssim&\mathcal{O}(0.10)& {\rm fb}\, ,\\
\Delta \sigma_{Zjj}^{\text{WBF}} &\lesssim&\mathcal{O}(0.003)& {\rm fb}\,
\end{array}
\renewcommand\arraystretch{0}
\end{equation}
following again the tool chain described in Sec.~\ref{sec:anatomysm}. As outlined in Section~\ref{sec:anatomysm}, the current experimental and theoretical uncertainties of the Standard Model cross sections are at the $\mathcal{O} (1-5\%)$ level. This makes it clear that it would be challenging to probe the $ZZZ$ vertex contributions from this model at the LHC in the absence of theoretical and experimental advances which would bring these uncertainties down by at least two orders of magnitude.

Since the overall impact of the fermionic scenario is slightly more promising than the scalar model we have discussed previously, we can raise the question of whether this scenario can be constrained at all using indirect collider measurements. While the LHC is limited by systematic uncertainties eventually, this situation is vastly improved for a future lepton collider. At such a machine we can expect measurements of electroweak diboson production to reach subpercentage-level precision, which offers an opportunity to see the imprint of vectorlike leptons in $ZZ$ measurements. We have calculated the size of the vectorlike lepton contribution for demonstrative ILC and CLIC setups, again using a calculation based on \textsc{FeynArts}, \textsc{FormCalc} and \textsc{LoopTools}, for a number of parameter points which are randomly distributed over the parameter space. The results are presented as a fraction of the Standard Model expectation in Fig.~\ref{fig:leptoncollider}.
The $Z$ boson pair production cross section in SM including SM-NLO electroweak effects is $\sigma(e^+e^-\to ZZ)\simeq 1.0~\text{pb}$ at $\sqrt{s}\simeq 250~\text{GeV}$ for on-shell $Z$ bosons in the on-shell scheme as implemented in~\cite{Hahn:1998yk,Hahn:2000kx} (these are also the choices that we adopt in the following)\footnote{See Refs.~\cite{Bohm:1986dn,Bohm:1986mz,Denner:1988nq,Denner:1988tv} for pioneering work.}. Experimental measurements of $ZZ$ production at LEP agree with this expectation, e.g. L3 report~\cite{Achard:2003hg}
\begin{equation}
\sigma/\sigma^{\text{SM}}=0.93\pm 0.08\text{(stat)}\pm 0.06 \text{(sys)}\,.
\end{equation}
L3 based their analysis GRC4F Monte Carlo~\cite{Fujimoto:1996wj}.\footnote{Special care is devoted in this analysis to handling of initial state radiation. GRC4F provides implementations based on the electron structure function and QED parton shower approaches based on Refs.~\cite{Kuraev:1985hb,Kurihara:1995mk}. The L3 $ZZ$ results~\cite{Achard:2003hg} include an associated 2\% uncertainty.}

Turning to the impact of new physics contributions on the full NLO electroweak cross section, we find that deviations $\lesssim 10\%$ for the total cross sections are possible, Fig.~\ref{fig:leptoncollider}, however, the bulk of parameter points that survive in our scan induce NLO deviations at the order of 3\%. These fall within the expected sensitivity of the ILC and CLIC proposals based on a purely statistical extrapolation (the shaded areas refer to the allowed regions of the different setups at different energies and luminosities). Note, that we do not include any systematic uncertainties in this comparison, which would imply a loss of sensitivity when exceeding 5\%. It is worthwhile to mention the ILC currently does not consider $ZZ$ production as a viable new physics candidate, and the main part of $ZZ$ production is to inform $WW$ measurements in data-driven approaches~\cite{Behnke:2013xla}.

Sizeable effects are also present for the case where the lightest mass eigenstate is too heavy to be pair-produced directly, however the modifications to the cross section decouple for $m_\text{lightest}\gg m_Z$ as can be expected from the general arguments of~\cite{Appelquist:1974tg}. This seems to be in contrast with our LHC findings and the decrease in sensitivity with $m_\text{lightest}$ is also slower than anticipated from the the discussion of the effective $ZZZ$ vertex, which deserves a comment. The reason behind this is that the effective $ZZZ$ and $\gamma ZZ$ interactions are not the driving force behind the corrections in Fig.~\ref{fig:leptoncollider}. The interactions of the new fermions with the $Z$ and $\gamma$ bosons induce modifications to the polarisation functions that enter in the definition of the renormalisation constants~(see e.g.~\cite{Denner:1991kt} for a comprehensive list). For instance, fixing gauge-kinetic terms and gauge boson masses on-shell, potential deviations from the SM lagrangian can become visible in the the interactions of gauge bosons with fermions~(see e.g.~\cite{Burgess:1993vc}). This is the basis of $S,T,U$ approach and implies that fermion gauge-boson interactions are sensitive to the presence of states that couple predominantly to gauge bosons. Such modifications can drive the relative change compared to the SM~\footnote{Similar observations have been made in the context of Higgs physics, see~\cite{Englert:2013tya,Craig:2013xia,Craig:2014una} and have motivated an extension of the oblique parameters to the Higgs boson~\cite{Gori:2013mia}.}. Our results can therefore be understood as $S,T,U$-like constraints at higher energies of ILC and CLIC compared to LEP. While off-shell production of $s$-channel vector bosons become statistically limited at such energies, the milder decrease of the $t$-channel $ZZ$ production therefore allows us to perform investigations along similar lines at future colliders at large statistics.

It should be stressed that these effects are correlated but understanding larger electroweak corrections in relation to anomalous $ZZZ$ interactions is a model-dependent statement and should therefore be taken with a grain of salt. However, while the challenging threshold results for the LHC do suggest that a plethora of new physics effects can still hide below the constraints of \cite{Sirunyan:2017zjc}, precision $ZZ$ measurements can be employed to constrain the models of $ZZZ$ interactions even when direct LHC constraints are loose.

\section{Summary and Conclusions}

Measurements of the electroweak sector of the Standard Model are well-motivated at the LHC as the high center-of-mass energy and luminosity allows us to test detailed predictions at unprecedented precision. Any deviations from the Standard Model expectation in the electroweak sector can also provide considerable insight into currently open problems by, for example, providing a source of sufficient $CP$-violation to explain the baryon asymmetry of the universe. In this paper we have investigated the use of simplified models to interpret measurements of the $ZZZ$ vertex, which provide a more realistic and consistent theoretical framework than the commonly employed anomalous coupling and effective field theory approaches when the new physics is close to the weak scale. Indeed given the EFT arguments below Eq.~\eqref{eq:d8ZZZops} the experimental constraints are not sufficient to argue that the NP generating the $ZZZ$ vertex is safely decoupling, a condition necessary for the use of form factors such as $f_{4,5}^Z$ or the effective field theory framework. We have discussed the anatomy of the vertex and how it arises in the Standard Model at one-loop, and argued that the minimal simplified scenario which allows for new contributions from new scalar states at one-loop is given by a $CP$-violating 2HDM. We have also considered a minimal simplified scenario where the vertex arises from threshold contributions from new fermion loops, given by a new generation of vectorlike leptons.

Our analysis suggests LHC measurements of the $ZZZ$ vertex are relatively insensitive to these scenarios once existing constraints are taken into account and electroweak thresholds are difficult to resolve from both the overall cross section contribution and QCD uncertainty perspective. At a future lepton collider a $ZZZ$ measurement could provide crucial new information, for example to confirm the vectorlike lepton nature of a new state discovered at the LHC. However, we find that the BSM effects at lepton colliders are dominated by general radiative rather than changes in the effective $ZZZ$ vertex. Accordingly, it needs to be stressed that such an observation would not directly point to new sources of $C$ or $CP$ violation along the lines described in this work, but would be a correlated effect of the presence of new states with interactions with SM fields.

As the fields that contribute to the $ZZZ$ vertex will manifest themselves predominantly as dimension 6 effects that can be case into $S,T,U$-like parameters, associated effects might be within the reach of future LHC measurements. While this lies beyond the scope of this work, Ref.~\cite{Frederix:2018nkq} provides an important technical development in scanning electroweak precision effects in an automised way, thus providing an avenue for a realistic signal and background collider study. We leave this for future work.

\acknowledgements
T.C. and M.J.D are supported by the Australian Research Council. C.E. is supported by the IPPP Associateship scheme and by the UK Science and Technology Facilities Council (STFC) under grant ST/P000746/1. K.N. has been supported by the College of Science \& Engineering of the University of Glasgow through a PhD Scholarship and by the NWO.

\bibliography{paper.bbl}

\end{document}